\documentclass[conference]{IEEEtran}
\IEEEoverridecommandlockouts
\usepackage{cite}
\usepackage{amsmath,amssymb,amsfonts}
\usepackage{algorithmic}
\usepackage{graphicx}
\usepackage{textcomp}
\usepackage{xcolor}
\def\BibTeX{{\rm B\kern-.05em{\sc i\kern-.025em b}\kern-.08em
    T\kern-.1667em\lower.7ex\hbox{E}\kern-.125emX}}

\usepackage{url}
\newcommand{\quotes}[1]{``#1''}
\DeclareMathOperator*{\argmin}{arg\,min}

\DeclareMathOperator{\atantwo}{atan2}
    
\begin{document}

\title{Where You Are Is What You Do: On Inferring Offline Activities From Location Data}

\author{\IEEEauthorblockN{Alameen Najjar}
\IEEEauthorblockA{\textit{Rakuten Institute of Technology} \\
Tokyo, Japan \\
alameen.najjar@rakuten.com}
\and
\IEEEauthorblockN{Kyle Mede}
\IEEEauthorblockA{\textit{Rakuten Institute of Technology} \\
Tokyo, Japan \\
kyle.mede@rakuten.com}
}

\maketitle

\begin{abstract}
In this paper we investigate the ability of modern machine learning algorithms in inferring basic offline activities,~e.g., shopping and dining, from location data. Using anonymized data of thousands of users of a prominent location-based social network, we empirically demonstrate that not only state-of-the-art machine learning excels at the task at hand~(F1 score$>$0.9) but also tabular models are among the best performers. The findings we report here not only fill an existing gap in the literature, but also highlight the potential risks of such capabilities given the ubiquity of location data and the high accessibility of tabular machine learning models.
\end{abstract}


\section{Introduction}
\label{sec1}

With the proliferation of social media, smartphones, Internet-of-things (IOT) devices and low Earth orbit (LEO) satellites, we live in a world where location data (Data with reference to a physical location) is ubiquitous. Recent estimates \cite{huang2020big} suggest that location data make up over 80\% of the data created on a daily basis. While location data can be and have been used for widely agreed on, positive applications — such as understanding the spread of infectious diseases \cite{jing2023using} — it can be easily misused. For example, misleading users of a navigation smartphone app about collecting and/or selling their data \cite{bbc}. 

It is well established that a user's location is indicative of the type of real-world, day-to-day activities, such as shopping and dining (Hereafter referred to as \quotes{offline activities}) they perform. Research has shown that basic offline activities can be inferred from GPS traces using both conventional statistical methods \cite{liao2005location, liao2005location2, liao2007extracting} and machine learning algorithms \cite{lane2011enabling, montini2014trip, martin2018graph, liu2022graph}. The same has been demonstrated using mobile phone data collected city-wide at the base station level \cite{noulas2013exploiting, liu2013annotating}. Other relevant sources of data, such as geotagged social media posts~\cite{lian2011collaborative, yang2014modeling}, WiFi signals \cite{zheng2011user}, and ride-hailing app data~\cite{hossain2021inferring} have also been successfully used to infer offline activities. In short, there is an abundance of evidence that points to the correlation between a person's location and the offline activity they are engaged in.

As with any type of location technology, opportunities exist for malicious targeting. For example, the same algorithm used in \cite{liu2022graph} to help patients with alcohol use disorder recover can be exploited to target users of a smartphone app most vulnerable to alcoholism. Such a risk is further amplified given the recent widespread availability of powerful machine learning algorithms \cite{he2021automl} that require little to no knowledge of statistics and/or algorithm design to configure and employ.

In this paper, we attempt to answer the following question: How well can modern machine learning algorithms infer user's offline activity given their location data? To this end, we empirically evaluate the performance of 6 models trained to infer 9 basic offline activities using \emph{anonymized} data collected over a period of 18 months from~$\approx$15k users of a prominent location-based social network active in 6 major cities spread over 4 different continents.

The findings we report in this paper not only fill an existing gap in the literature, but also highlight the potential risks of applying machine learning to location data in a time where powerful machine learning models are easily accessible. The following is a summary of our most interesting findings:

\begin{itemize}
    \item Our experiments show that modern machine learning algorithms are well capable of inferring basic offline activities with the best performing model achieving an average Macro-F1 score of over 0.9.
    \item We also found that \quotes{Nightlife} is the most and \quotes{At home} is the least challenging activities to infer on average.
    \item Finally, we found that tabular models that require minimal machine learning knowledge to configure and limited resources to run not only excel at the task at hand but could also outperform sophisticated models trained end-to-end.
\end{itemize}

The remainder of this paper is organized as follows. Previous relevant works are briefly overviewed in Section \ref{sec2}. The methodology we follow to infer offline activities from location data is described in Section \ref{sec3}. The results of our extensive empirical analysis are given in Section \ref{sec4}. And finally the paper is summarized in Section \ref{sec5}.

\section{Previous Works}
\label{sec2}
The existing literature on inferring offline activities is vast, and it is beyond the scope of this paper to review it in its entirety. Instead we have organized collections of representative works into 4 categories based on the data used; those being: mobility diaries, Call Detail Records, social check-ins and others.

The bulk of the literature use self-reported mobility diaries recorded using specialized hardware and/or software \cite{liao2005location, liao2005location2, liao2007extracting, lane2011enabling, montini2014trip, martin2018graph, liu2022graph}. The data used is dense however limited in terms of number of subjects, temporal span and spatial coverage. Early works \cite{liao2005location, liao2005location2, liao2007extracting} used Conditional Random Fields (CRFs) and Relational Markov Networks (RMNs) to infer basic offline activities from GPS traces of a few subjects. In the same vain, 
 \cite{lane2011enabling} proposed a personalized framework that leverages similarities among users to infer offline activities from GPS traces of 50 subjects. In \cite{montini2014trip}, Random Forests are used to infer the purpose of trips~(High-level offline activities) of GPS traces of 156 subjects collected over a one week period around Zurich, Switzerland. Similarly in \cite{martin2018graph} graph convolutional neural networks (GCNs) \cite{kipf2016semi} are used to classify GPS traces of 139 subjects into 5 offline activities. GCNs are also used recently in~\cite{liu2022graph} to infer 8 offline activities in GPS diaries collected and labeled by~167 subjects.

A second subset of the literature \cite{noulas2013exploiting, liu2013annotating} infer offline activities at the base station level from Call Detail Records~(CDR) provided by mobile network operators. In \cite{noulas2013exploiting}, conventional machine learning algorithms are used to infer 8 offline activities from CDR data collected in Barcelona and Madrid. Similarly \cite{liu2013annotating} uses an ensemble of models to classify CDR data of 80 users into 5 basic offline activities.

A third subset of the literature \cite{lian2011collaborative, yang2014modeling} use social check-in data to infer user's offline activities at the Point Of Interest (POI) level. In \cite{lian2011collaborative}, CRFs combined with unsupervised clustering is used to infer 7 offline activities from DianPing\footnote{\url{https://www.dianping.com/}} check-in data of 83 users collected in Beijing, China. Similarly~\cite{yang2014modeling} uses non-zero matrix factorization to infer 9 offline activities from Foursquare\footnote{\url{https://foursquare.com/}} check-in data of $\approx$2000 users collected in Tokyo and New York city.

The fourth and final subset of the literature are works that use data sources other than those listed above. For example,~\cite{zheng2011user} infers 8 offline activities from WiFi traces of 13 subjects moving around a university campus, \cite{song2013collaborative, cui2019tweets} infer offline activities from microblogs, and \cite{hossain2021inferring} infers 13 high-level activities from ride-hailing app data collected around the city of Toronto, Canada.

Data wise, our work mostly resembles that of \cite{lian2011collaborative, yang2014modeling}. However, we are not aware of any previous work that attempted to infer offline activities at the scale we report here~(6 cities, $\approx$15k users, and 6 models). Finally, it is worth mentioning that works, such as~\cite{ye2013s, liao2018predicting, to2019traveler} that \quotes{predict} future offline activities using data similar to ours are beyond the scope of this survey as we are interested in inferring the user's current rather than future activities.

\section{Methodology}
\label{sec3}
In this section we explain the methodology we follow to infer offline activities from location data.

\subsection{Preliminaries}
\bigbreak

\noindent{\textbf{Definition 1} (Check-in record)}. \emph{A check-in record is a tuple~$\langle u, l, t\rangle$ indicating that user $u$ is present at location $l$ at time $t$}, where $l$ is the location of a uniquely identified POI. \bigbreak

\noindent{\textbf{Definition 2} (Offline activity)}. \emph{The offline activity associated with a given check-in record is the category of the POI at which the check-in takes places}. For example, \quotes{Dining} is the activity associated with \quotes{Food} POIs, and \quotes{At home} is the activity associated with \quotes{Residence} POIs. It is worth noting that using POI categories as activities is a common practice as it has been done previously in~\cite{lian2011collaborative, ye2013s, noulas2013exploiting, song2013collaborative, yang2014modeling, liao2018predicting, cui2019tweets}.  \bigbreak

\noindent{\textbf{Problem} (Activity inference)}. Given a check-in record $\langle u, l, t\rangle$ of user $u$, the objective is to infer the activity user $u$ is engaged in at time $t$ and location $l$. Let $\mathcal{X} = \{x_1, x_2, \cdot\cdot\cdot, x_n\}$ represent the set of check-in records and $\mathcal{Y} = \{y_1, y_2, \cdot\cdot\cdot, y_m\}$ represent the set of activities, the goal is to find a function $f(\cdot)$ that assigns each record in $\mathcal{X}$ with one activity in $\mathcal{Y}$ while satisfying the following condition:

\begin{equation}
    \argmin_{f \in \mathcal{F}} \lVert f(x_i) - y_i \rVert, f(x_i) \in \mathcal{Y}, y_i \in \mathcal{Y},
\end{equation}

\noindent where $y_i$ is the true label (Activity) associated with $x_i$, $\lVert \cdot \rVert$ is an evaluation operator (That evaluates to 0 when $f(x_i)=y_i$ and 1 otherwise), and $\mathcal{F}$ is the hypothetical space of the task at hand. 

\subsection{Enrichment}
To account for meaningful contextual information, we enrich check-in records with two sets of features as follows.

\emph{Relative location}. By relative location we mean location with respect to the center of the city. More specifically, 1)~distance to city center, and 2) bearing angle with respect to city center. Our intuition behind including these two features is based on the assumption that POIs of the same category are more likely to share similar spatial distribution patterns with respect to the center of the city. For example, camping grounds are more likely to be found in the periphery of the city.

Distance to city center ($\delta$) is calculated using the Haversine formula as follows:

\begin{equation}
    a = \sin^2 (\frac{\Delta_{\phi}}{2}) + \cos \phi_1 \cdot \cos \phi_2 \cdot \sin^2 (\frac{\Delta_{\lambda}}{2}),
\end{equation}

\begin{equation}
    c = 2 \cdot \atantwo(\sqrt{a}, \sqrt{1 - a} ),
\end{equation}

\begin{equation}
    \delta = R \cdot c,
\end{equation}

\noindent where $\phi$ is latitude, $\lambda$ is longitude, $\Delta_{\phi}$ is the difference between two latitudes, $\Delta_{\lambda}$ is the difference between two longitudes, and $R$ is Earth's radius. 

On the other hand, bearing angle with respect to city center~($\theta$) is calculated such as:

\begin{equation}
    A = \sin \Delta_{\lambda} \cdot \cos \phi_2,
\end{equation}

\begin{equation}
    B = \cos \phi_1 \cdot \sin \phi_2 - \sin \phi_1 \cdot \cos \phi_2 \cdot \cos \Delta_{\lambda},
\end{equation}

\begin{equation}
    \theta = \atantwo(A , B),
\end{equation}

\noindent where $\phi$ is latitude, $\lambda$ is longitude, and $\Delta_{\lambda}$ is the difference between two longitudes.

\emph{Grid statistics}. We extract three POI-related statistics calculated at the grid cell level, namely POI count, unique user count and check-in count. Our intuition behind including these features is to help the model capture meaningful patterns on the spatial distribution of different POI categories around the city. For example, the number of POIs around residential areas is likely to be less than that around commercial areas. Or grid cells with high check-in activity are less likely to be in a residential area, and so forth. Each of the features can be described as follows:

\begin{equation}
    \psi_{c} = \sum_{i \in M} i, c \in N,
\end{equation}

\noindent where $\psi_c$ is the extracted feature at cell $c$, $i$ is the $i$-th POI/user/check-in per cell $c$, $M$ is the set of unique POIs/users/check-in per cell $c$, and $N$ is the set of all cells in the target city.

\emph{Multi-scale \& multi-grid feature extraction}. To account for a richer feature set, we extract the aforementioned features at multiple spatial scales using multiple hierarchical grids.

\subsection{Encoding \& Classification}
After enrichment, each check-in is represented with a vector $v \in \mathbb{R}^d$ made from concatenating the check-in attributes (User ID, aggregated location, and timestamp-related attributes) and the enriched features, such that:

\begin{equation}
    v = (u, t, l, \delta, \phi, \psi_1, \cdot \cdot \cdot, \psi_k) \in \mathbb{R}^d,
\end{equation}

\noindent where $u$ is the user ID, $t$ is the timestamp, $l$ is the aggregated location, $\delta$ and $\theta$ are the relative location features, $\psi$ is the grid statistics, and $k$ is the number of scales and/or grids used in the check-in enrichment step. 

The representation obtained in Equation 9 is used next for classification. To this end, we experimented with 6 classifiers organized into 3 groups: conventional classifiers including both $k$-Nearest Neighbours ($k$-NN) \cite{fix1989discriminatory} and Support Vector Machines (SVM) \cite{cortes1995support}. Tabular classifiers including Extreme Gradient Boosting (XGB) \cite{chen2016xgboost} and TabNet \cite{arik2021tabnet}. And Multilayer Perceptrons \cite{rosenblatt1958perceptron} implemented in two flavors: vanilla/plain~(MLP) and regularized (rMLP). See Appendix \ref{apxa} for model implementation details.

\section{Results}
\label{sec4}
In this section we present the validation results of inferring offline activities from location data.

\begin{figure}[ht]
  \centering
  \includegraphics[width=0.9\linewidth]{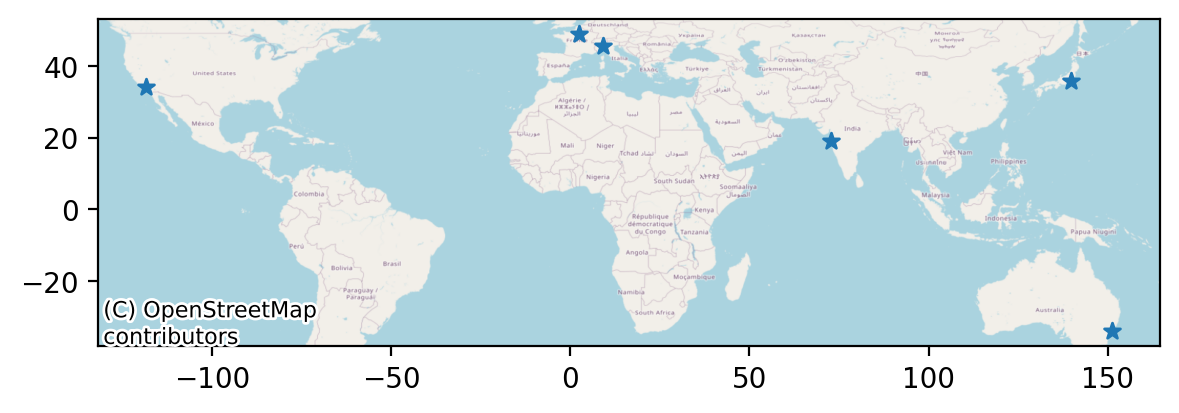}
  \caption{Map of the target cities: Los Angeles, Tokyo, Mumbai, Sydney, Paris and Milan.}
  \label{fig1}
\end{figure}

\begin{figure}[ht]
  \centering
  \includegraphics[width=0.9\linewidth]{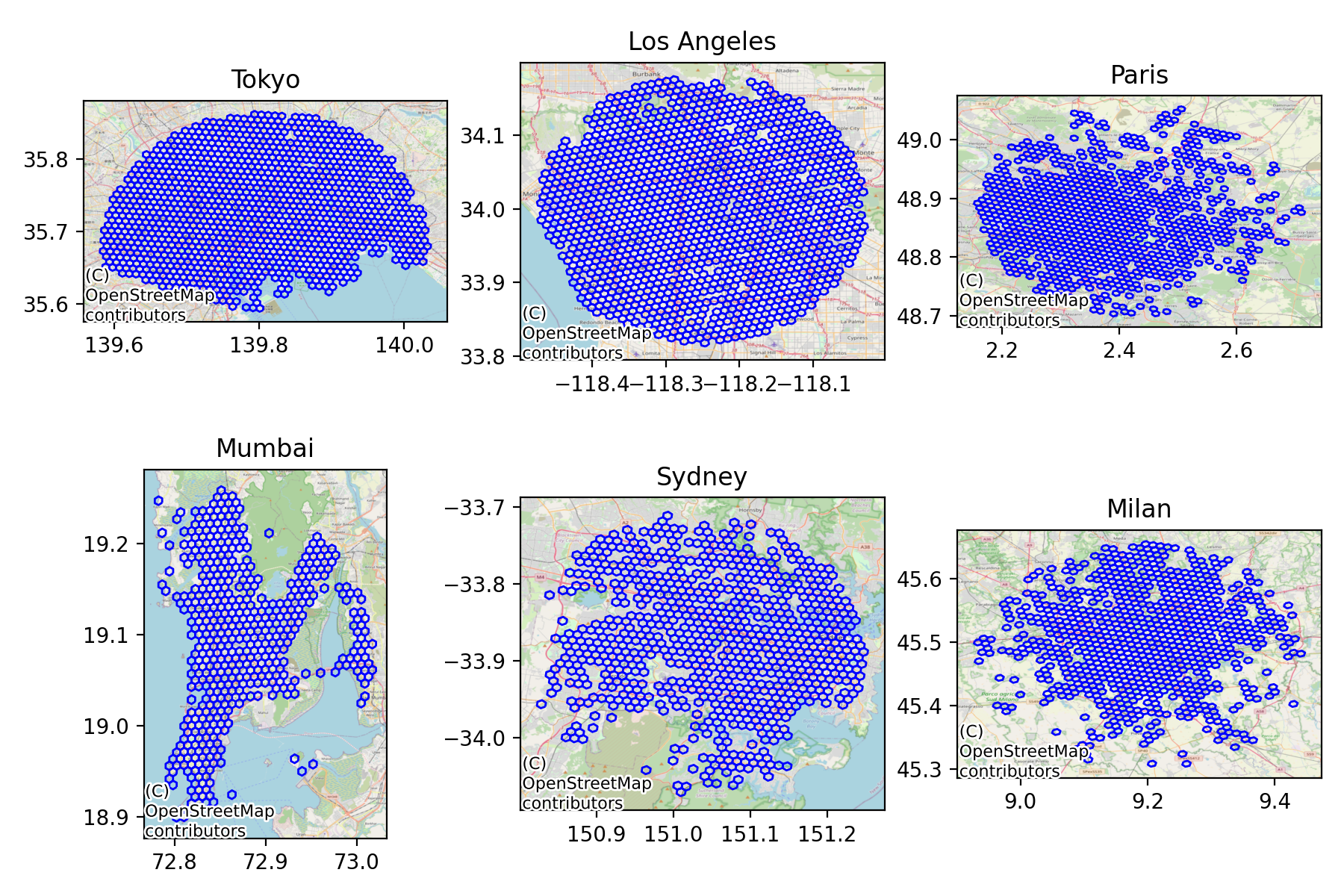}
  \caption{Data coverage per target city. Missing polygons indicate data gaps.}
  \label{fig2}
\end{figure}

\subsection{Data}
We used the Foursquare dataset \cite{yang2016participatory} since it is one of the most widely used check-in dataset in the research community. The dataset consists of over 33M check-ins made by over~266k users at over 3.6M venues in 415 cities worldwide over a period of 18 months. 

\emph{For privacy concerns we opted out of using the user's actual location. Instead we aggregated location using a grid, such as Uber H3\footnote{\url{https://github.com/uber/h3}} and Geohash\footnote{\url{http://geohash.org/}}. In other words, check-ins within the same grid cell are indistinguishable to one another from the location point of view}. Venue names and user names, on the other hand, are already anonymized by the source therefore we used them as they are.

Next, we kept check-ins that belong to 6 major cities spanning a wide range of latitudes and longitudes, and representing~6 sub-regions as defined by the United Nations Geoscheme\footnote{\url{https://en.wikipedia.org/wiki/United_Nations_geoscheme}}. Each check-in is assigned a city based on its distance to the city center's coordinates as provided in the dataset. See Figure \ref{fig1} for a map of the selected cities. Target cities have good data coverage with little data gaps. See Figure~\ref{fig2} for a visualization of data coverage.

Finally, we assigned each venue a category out of nine parent categories as defined by the Foursquare API\footnote{\url{foursquare-categories.herokuapp.com/}}. In other words, each venue is assigned one of the following categories: \quotes{Arts \& Entertainment,} \quotes{College \& University,} \quotes{Food,} \quotes{Nightlife Spot,} \quotes{Outdoors \& Recreation,} \quotes{Professional \& Other Places,} \quotes{Residence,} \quotes{Shop \& Service,} and \quotes{Travel \& Transport}. These categories are the target variables we aim to infer given a check-in.

Cities are treated as separate datasets and analyzed independently as reported in the following. See Table \ref{tab1} for a summary of the statistics of the final datasets.

\begin{table}
\centering
  \caption{Summary of the generated datasets.}
  \label{tab1}
  \begin{tabular}{|l|lll|}
    \hline
    City & Check-ins & Venues & Users\\
    \hline
    Los Angeles & 97274  & 18685 & 2476\\
    Tokyo       & 708863 & 64761 & 8360\\
    Mumbai      & 25248  & 6088  & 525\\
    Sydney      & 31934  & 7739  & 733\\
    Paris       & 59816  & 13603 & 1973\\
    Milan       & 40641  & 8642  & 897\\
    \hline
    Total       & 963776 & 119518  & 14964 \\
  \hline
\end{tabular}
\end{table}







\subsection{Model Comparison}
In Table \ref{tab2} we report the validation loss of all models on all 6 datasets. See Appendix \ref{apxa} for experiment implementation details.

\begin{table*}
\centering
  \caption{Validation loss (Log loss $\pm$ standard deviation) obtained by different models on all datasets.}
  \label{tab2}
  \begin{tabular}{|l|lllllll|}
    \hline
           & Mumbai & Sydney & Milan & Paris & Los Angeles & Tokyo & Average\\
    \hline
    $k$-NN & 1.968$\pm$0.023 & 1.891$\pm$0.043 & 1.888$\pm$0.019 & 1.963$\pm$0.04 & 1.914$\pm$0.022 & 1.596$\pm$0.011 & 1.870$\pm$0.003 \\
    SVM & 1.742$\pm$0.009 & 1.725$\pm$0.002 & 1.787$\pm$0.002 & 1.882$\pm$0.004 & 1.894$\pm$0.002 & 1.556$\pm$0.004 & 1.764$\pm$0.004 \\
    XGB & \textbf{0.868$\pm$0.011} & \textbf{0.832$\pm$0.003} & \textbf{0.709$\pm$0.009} & \textbf{0.811$\pm$0.01} & \textbf{0.755$\pm$0.005} & \textbf{0.598$\pm$0.001} & \textbf{0.762$\pm$0.006} \\
    TabNet & 0.951$\pm$0.062 & 1.099$\pm$0.006 & 1.007$\pm$0.049 & 1.003$\pm$0.007 & \underline{0.895$\pm$0.031} & 0.813$\pm$0.016 & 0.961$\pm$0.028 \\
    MLP & 1.110$\pm$0.009 & 1.191$\pm$0.021 & 1.047$\pm$0.025 & 1.125$\pm$0.005 & 0.94$\pm$0.011 & \underline{0.72$\pm$0.003} & 1.022$\pm$0.012 \\
    rMLP & \underline{0.937$\pm$0.01} & \underline{0.994$\pm$0.011} & \underline{0.869$\pm$0.002} & \underline{0.931$\pm$0.004} & 0.924$\pm$0.024 & 0.757$\pm$0.004 & \underline{0.902$\pm$0.009} \\
    \hline
  \end{tabular}
\end{table*}

Key takeaways can be summarized as follows. XGB dominates all models on every single dataset. On average, XGB reduces the naive model's ($k$-NN) loss by $59\%$. Interestingly, XGB's performance is on average~$20.9\%$ better than that of TabNet which is Google's deep-learning  model designed for tabular data; although inline with previous studies \cite{kadra2021well, shwartz2021tabular}.

Second in place is rMLP which cuts the naive model's loss by~$51.7\%$. Still rMLP performs $18\%$ worse than XGB. Regularization boosts plain MLPs by an average of $11.7\%$ which we take as a demonstration of the potential regularization has for MLPs. It is worth noting that rMLP outperforms TabNet on all datasets except Los Angeles and on average it provides a $6\%$ performance boost over TabNet. This finding is inline with the results in \cite{kadra2021well}.

Moreover, TabNet performs better than plain MLPs on all datasets except Tokyo. On average TabNet outperforms plain MLPs by $\approx6\%$.

Given the above results, we consider XGB the winning model and therefore we use it hereafter in our experiments. Next, we empirically attempt to understand how individual features contribute to the model's performance.

\begin{figure}[ht]
  \centering
  \includegraphics[width=0.9\linewidth]{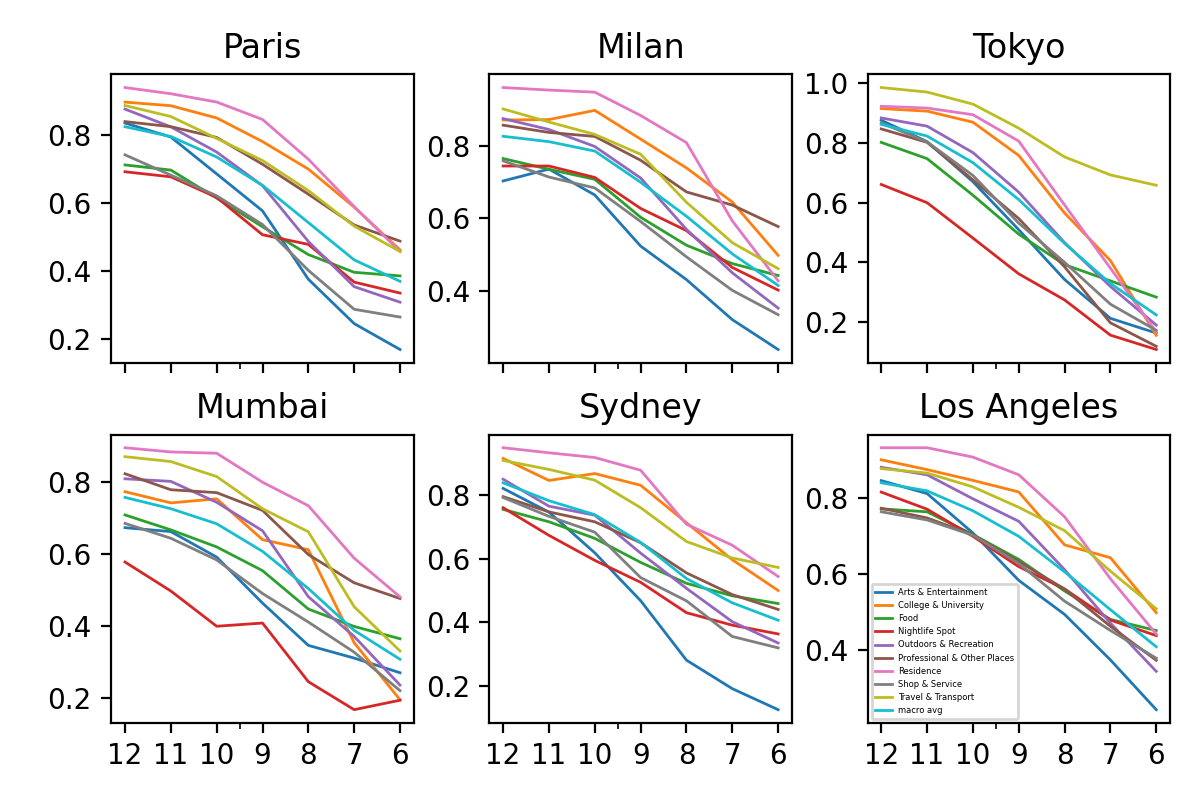}
  \caption{Location granularity and performance: Classification performance (Y axis) plotted against decreasing grid resolution (X Axis).}
  \label{fig3}
\end{figure}

\begin{figure}[ht]
  \centering
  \includegraphics[width=0.9\linewidth]{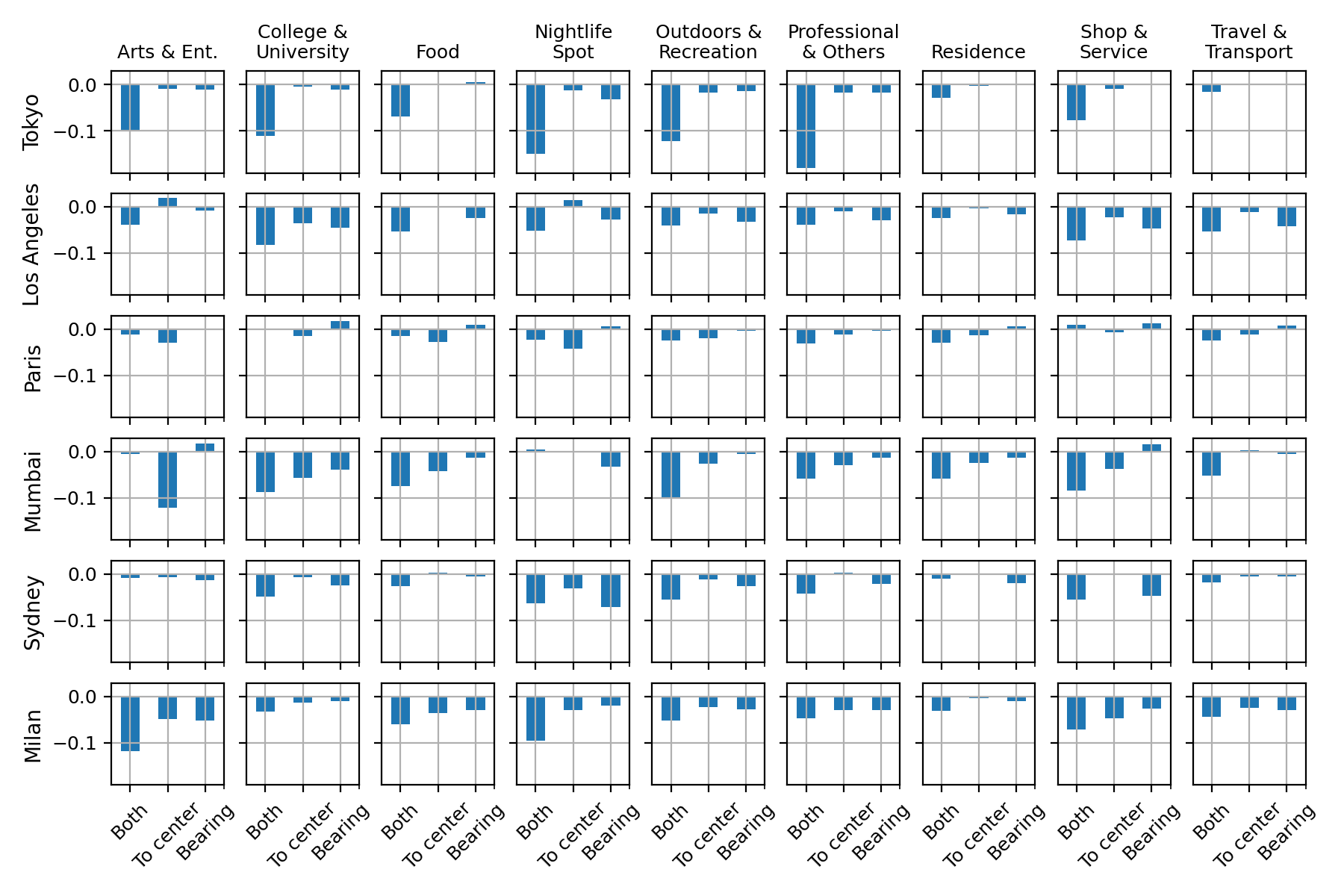}
  \caption{Relative location and performance: Percentage of performance change (Y axis) resulted from excluding different relative-location features (X axis).}
  \label{fig4}
\end{figure}

\begin{figure}[ht]
  \centering
  \includegraphics[width=0.9\linewidth]{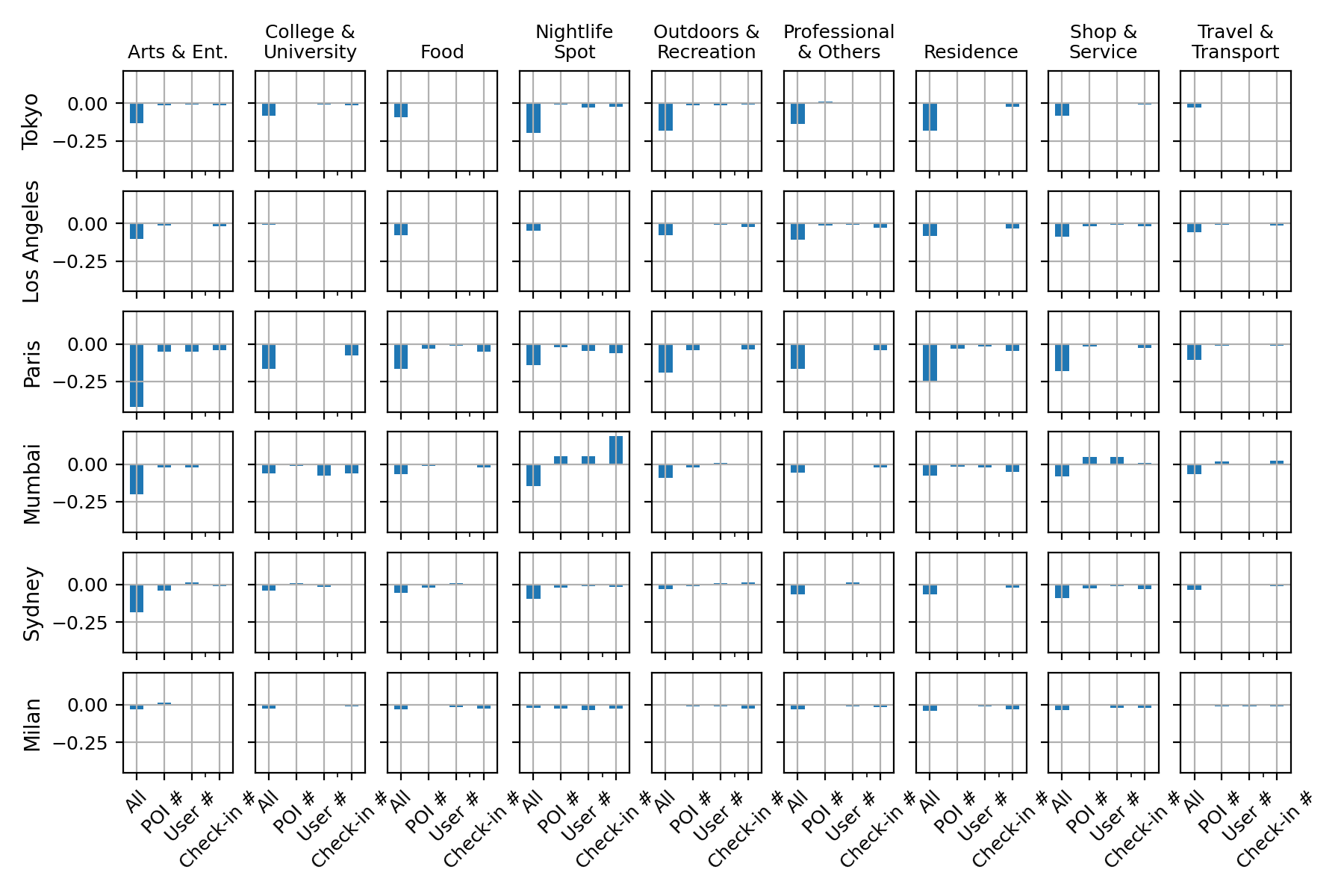}
  \caption{Grid statistics and performance: Percentage of performance change (Y axis) resulted from excluding different grid statistics (X axis).}
  \label{fig5}
\end{figure}

\subsection{Ablation Study}
In order to understand how different features contribute to the model's performance in the following we present the results of a series of ablation experiments we conducted.

Starting with location, in Figure \ref{fig3} we plotted the per-category classification performance (Macro-F1) against a decreasing grid resolution. The general trend indicates that higher grid resolution yields better performance. City wise, performance degrades by an average of 57\% when grid resolution goes from highest to lowest with Milan and Tokyo being the least and most impacted cities. Category wise on the other hand, \quotes{Travel \& Transport} and \quotes{Arts~\& Entertainment} are the least and most impacted.

While the above results indicate that location data play an essential role in the model's performance what is more interesting is the observation that even when features are extracted using the lowest resolution grid the model is still able to correctly infer the user activity up to 36\% of the time. This is indicative of the importance of non-location features.

Moving on to relative location, in Figure \ref{fig4}, we plotted the change in performance obtained when different relative-location features are removed. Different cities and/or categories are impacted differently by relative location. However, on average excluding relative information degrades performance by 2\% to 9\% with Paris and Tokyo being the least and most impacted cities. Moreover, \quotes{Residence} and \quotes{Outdoors \& Recreation} are the least and most impacted categories. Individually, \quotes{Nightlife Spot} and \quotes{Arts \& Entertainment} benefit the most from bearing angle and distance-to-center, respectively. 

Next we evaluated grid statistics in Figure \ref{fig5}. Removing grid statistics degrades performance by and average of 9.6\% with Milan and Paris being the least and most impacted cities. Category wise, \quotes{Travel \& Transport} and \quotes{Arts \& Entertainment} are the least and most impacted categories. Different statistics contribute differently to the model's performance with check-in count being the most important among the three. Followed by user count and finally POI count with a very small margin in between.

The obtained results demonstrate that both \emph{relative information} and \emph{grid statistics} are important to the model's performance and thus confirm the assumptions we made earlier.

\begin{figure}[ht]
  \centering
  \includegraphics[width=0.9\linewidth]{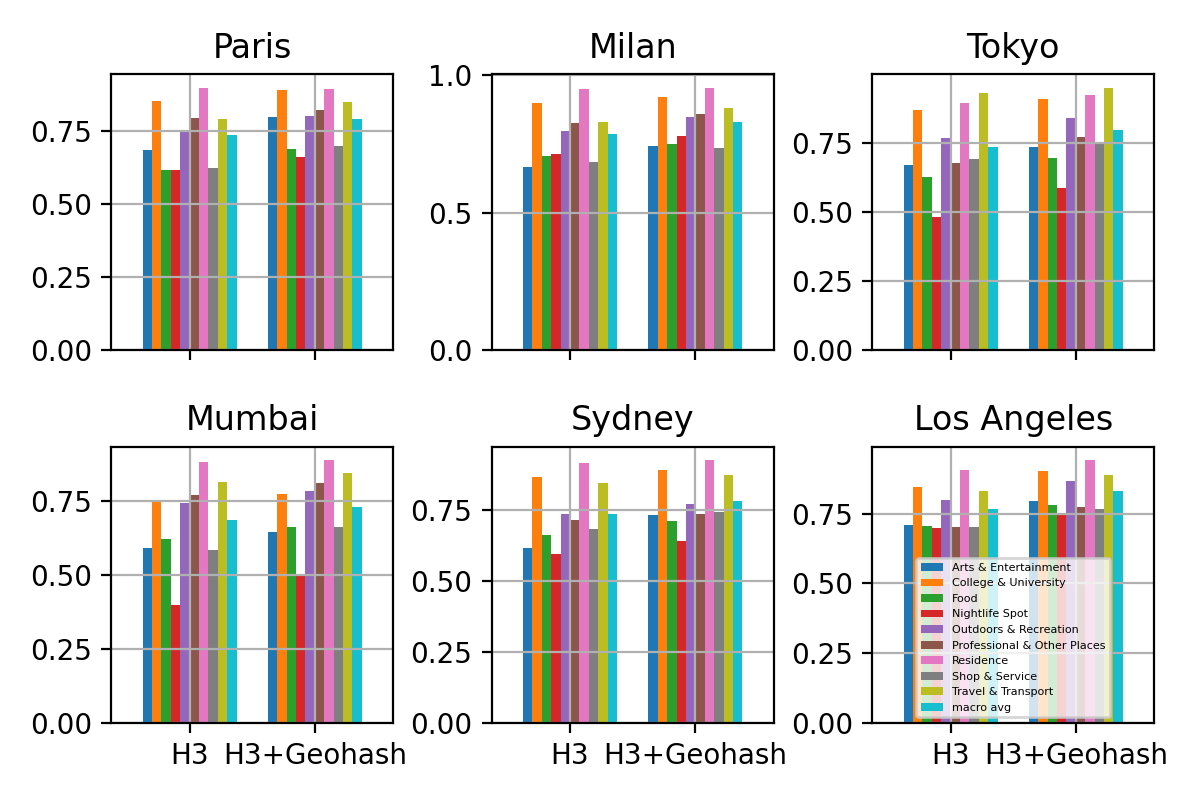}
  \caption{Multi-grid feature extraction and performance: Per-category classification performance (Y axis) plotted against number of grids (X axis).}
  \label{fig6}
\end{figure}

In Figure \ref{fig6} we compared the model's performance when features are extracted using one (H3) versus two (H3 and Geohash) grids. Using two grids instead of one boosts performance by an average of 7\%. Tokyo and Los Angeles benefit the most while Milan and Sydney benefit the least. Category wise, \quotes{Nightlife Spot} and \quotes{Residence} are the most and least impacted categories. In fact, \quotes{Residence} in Tokyo is negatively impacted when two grids are used instead of one.

\begin{figure}[ht]
  \centering
  \includegraphics[width=0.9\linewidth]{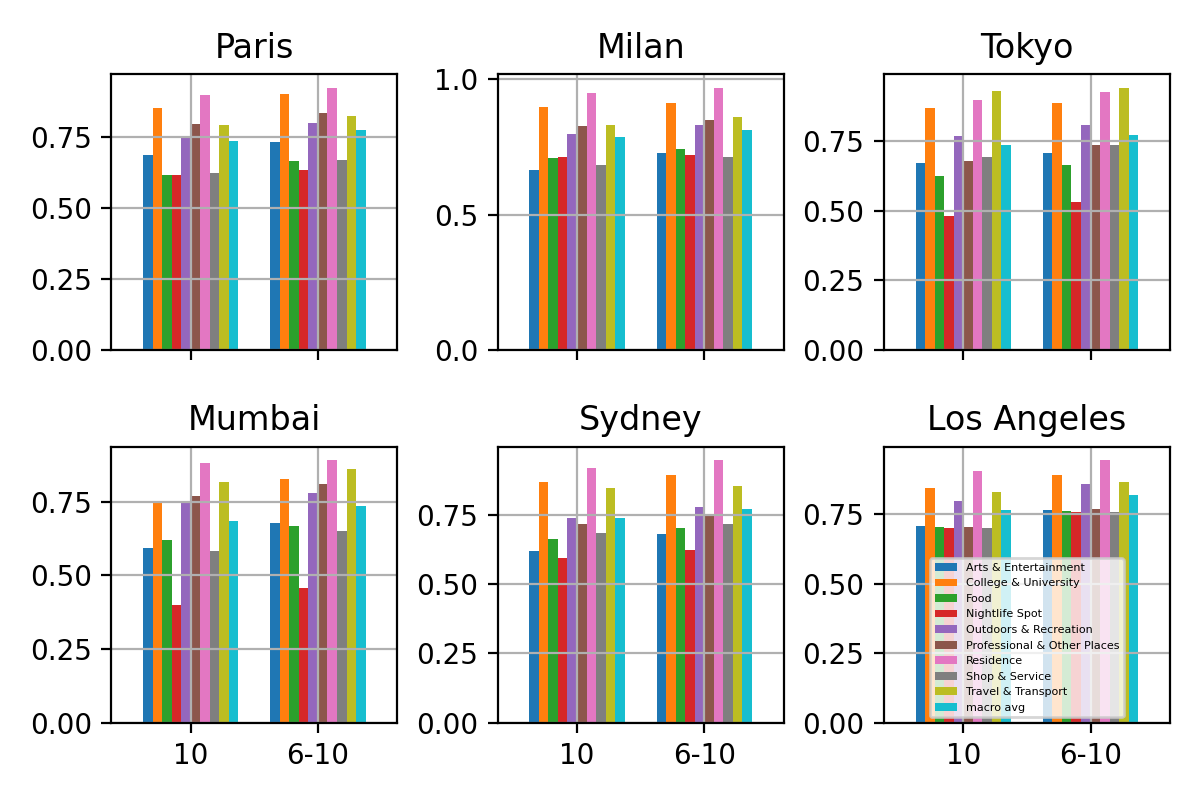}
  \caption{Multi-scale feature extraction and performance: Per-category classification performance (Y axis) plotted against number of grid scales (X axis).}
  \label{fig7}
\end{figure}

Finally, in Figure \ref{fig7} we studied the impact multi-scale features have on the model's performance. It is worth noting that both models (Single-/multi-scale) have the same location granularity. The obtained results show that all categories across all cities benefit from multi-scale feature extraction. On average performance is boosted by 5.4\%. Mumbai and Milan are the most and least impacted cities. On the other hand, category wise, \quotes{Arts \& Entertainment} and \quotes{Residence} are the most and least impacted categories.

\begin{figure}[ht]
  \centering
  \includegraphics[width=0.9\linewidth]{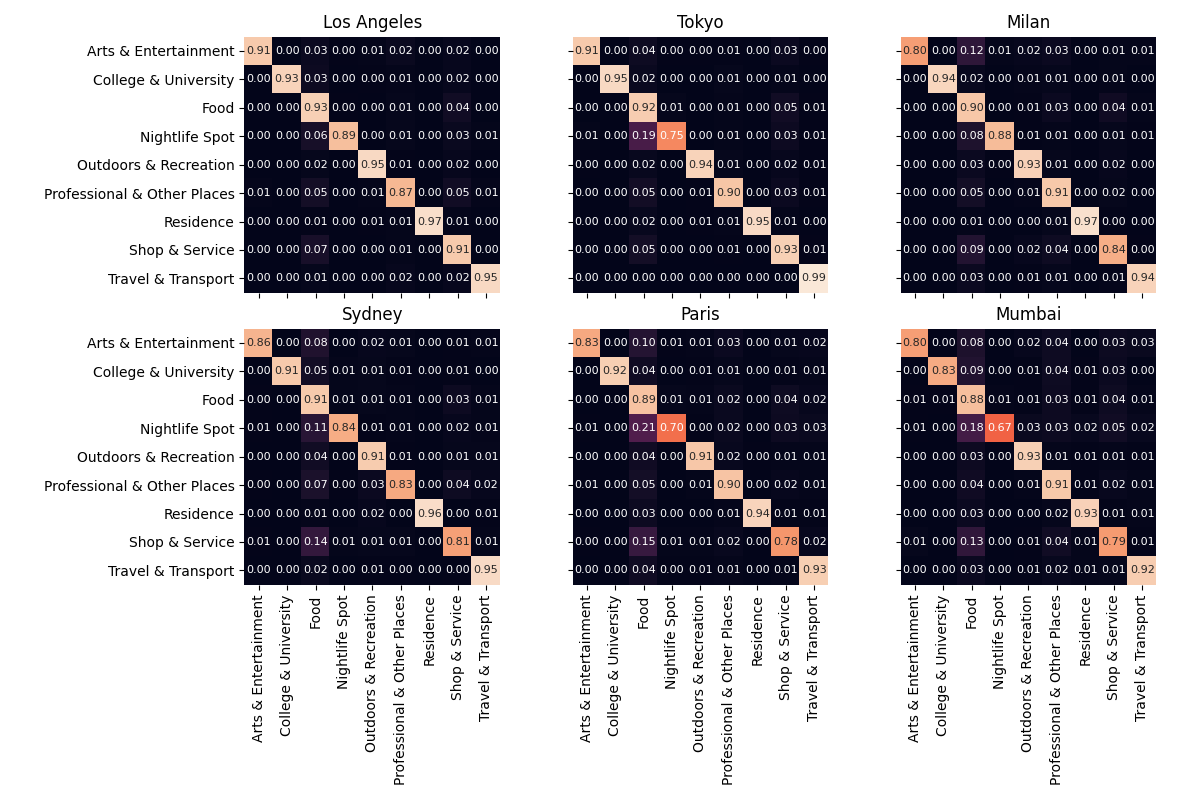}
  \caption{Best model evaluation (2): Normalized confusion matrix for all cities.}
  \label{fig12}
\end{figure}

In the following we build upon the insights we gained from the results above to evaluate our final model.

\subsection{Best Model Evaluation}
For a final evaluation we retrained the winning model using all features extracted using 2 grids at 7 different scales. Evaluation results of this model are reported in Table \ref{tab3}. 

On average and over all cities, the best model achieves a Macro-F1 score of 0.904. City-wise, the model's performance is comparable with Mumbai and Los Angeles being on the opposite ends of performance. The same observation holds true at the category level with \quotes{Nightlife Spot} and \quotes{Residence} being the most and least challenging categories. In general, the model performance is consistent across city and category with a standard deviation of $2.4\%$ and $4.6\%$, respectively.

To better understand how the model performs across category we plotted the confusion matrix in Figure \ref{fig12}. While the matrix shows mostly clear separation between categories, it is clear that that \quotes{Nightlife Spot} check-ins are largely misclassified as \quotes{Food} in almost all cities. The same behaviour is observed with \quotes{Shop \& Service} however to a lesser degree.

\begin{table*}
\centering
  \caption{Best model evaluation (1): Per-category Macro-F1 score for all cities.}
  \label{tab3}
  \begin{tabular}{|l|llllll|l|}
    \hline
           & Mumbai & Sydney & Milan & Paris & Los Angeles & Tokyo & Average\\
    \hline
    Arts \& Entertainment        & 0.849 & 0.883 & 0.861 & 0.889 & 0.932 & 0.934 & 0.891 \\
    College \& University        & 0.885 & 0.935 & 0.956 & 0.952 & 0.949 & 0.965 & 0.94 \\
    Food                         & 0.834 & 0.859 & 0.853 & 0.796 & 0.91  & 0.89  & 0.857 \\
    Nightlife Spot               & 0.767 & 0.882 & 0.915 & 0.776 & 0.92  & 0.807 & 0.844 \\
    Outdoors \& Recreation       & 0.929 & 0.896 & 0.938 & 0.927 & 0.958 & 0.948 & 0.933 \\
    Professional \& Other Places & 0.9   & 0.869 & 0.908 & 0.895 & 0.878 & 0.916 & 0.894 \\
    Residence                    & 0.937 & 0.976 & 0.982 & 0.956 & 0.979 & 0.974 & 0.967 \\
    Shop \& Service              & 0.812 & 0.839 & 0.847 & 0.807 & 0.9   & 0.925 & 0.855 \\
    Travel \& Transport          & 0.933 & 0.955 & 0.954 & 0.934 & 0.96  & 0.993 & 0.955 \\
    \hline
    Average                      & 0.872 & 0.899 & 0.913 & 0.881 & 0.932 & 0.928 & 0.904 \\
    \hline
  \end{tabular}
\end{table*}

\begin{figure*}[ht]
  \centering
  \includegraphics[width=0.8\linewidth]{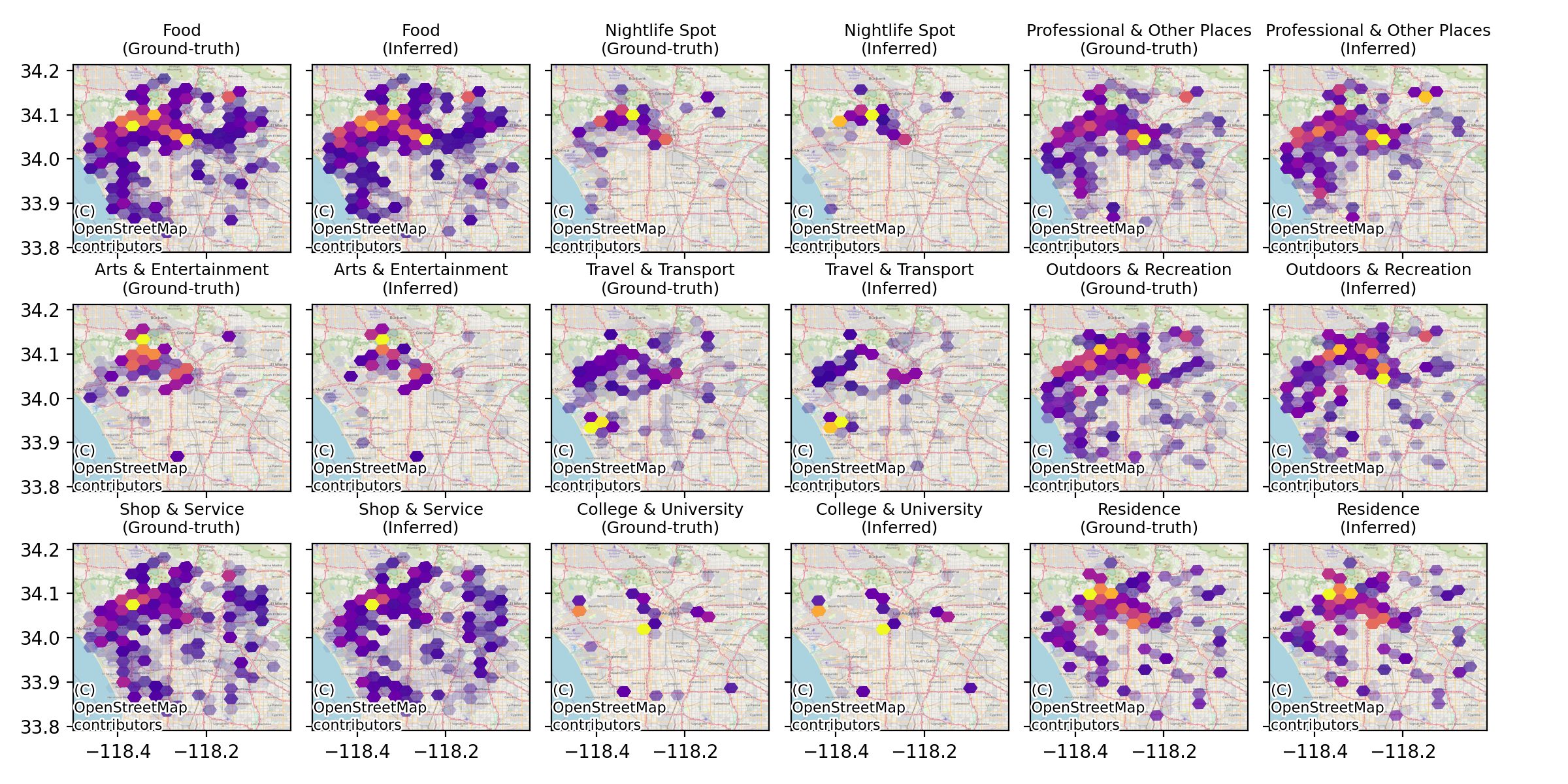}
  \caption{Best model evaluation (3): Aggregated inferences compared to ground-truth data for the city of Los Angeles.}
  \label{fig13}
\end{figure*}

Finally, for a subjective evaluation we mapped in Figure~\ref{fig13} the aggregated inferences made by the best model next to the ground-truth data for the city of Los Angeles. The visualizations clearly indicate that the inferred maps preserve to a high degree the spatial distribution of the data for the majority of classes. It is worth noting that we obtained similar results on the other cities. And that our choice of Los Angeles is based on both data coverage and model performance on the test data.

\subsection{Summary}
The following is a summary of the reported results. First, XGB is the best performing model by a large margin followed by regularized MLPs and TabNet, respectively. Second, the winning model is well capable of inferring offline activities with an average Macro-F1 score of 0.904. Third, performance is consistent across city and activity with a standard deviation value of 2.4\% and 4.6\%, respectively. Fourth, we found that \quotes{Nightlife} is the most and \quotes{At home} is the least challenging offline activities to infer with the winning model achieving an average Macro-F1 score of 0.844 and 0.967, respectively. Finally, the ablation study demonstrated that location granularity, relative location and grid statistics each on its own plays a significant role in the model's performance.

\section{Discussion}
\label{sec5}

Thanks to recent software and hardware advances, we live in a world where location data is ubiquitous. Previous research has well demonstrated that location is an effective proxy for the type of activity a person is engaged in in the real world.

In this paper, we attempted to answer the following question: How well can modern machine learning algorithms infer offline activities from location data? To this end, we empirically evaluated the performance of 6 models trained to infer 9 basic offline activities using \emph{anonymized} data collected from~$\approx$15k Foursquare users active in 6 major cities spread across 4 continents. Our experiments show that not only modern machine learning algorithms are well capable of inferring basic offline activities (Macro-F1$>$0.9) given location data, but also tabular models which require minimal knowledge to configure and limited resources to run are among the best performers. 

As with the majority of studies, ours is subject to limitations. First, social check-in data is subject to bias since it comes from public social media posts shared willingly by individuals who may be less concerned with privacy and not representative of the whole population. Second, it is not always the case that activities match the category of the POI at which a person checks in. For example, checking in at a \quotes{Residence} POI does not always mean engaging in \quotes{At home} activity. It could also mean \quotes{Work} if the person works from home. Similar arguments could be made about other POI categories. Therefore, the empirical findings reported in this work should be seen in the light of such limitations.

Finally, it is worth mentioning that some of the capabilities demonstrated in this paper could be easily misused by untrustworthy parties. For example, assigning users of a smartphone app negative or unwanted labels, such as \quotes{Unhealthy} or \quotes{Overweight} inferred from their location history. This concern is further amplified given the ubiquity of location data and the recent widespread of accessible yet powerful machine learning models.

\bibliographystyle{IEEEtran}
\bibliography{bibs}

\appendices
\section{Implementation Details}
\label{apxa}

\textbf{Data}. We used $80\%$ of each dataset for training and validation, and $20\%$ for testing. The training/validation subset is used for hyper-parameters tuning following a 3-fold cross validation evaluation scheme.

\textbf{Spatial grid}. We used resolution 10 Uber H3 grid for feature extraction. For multi-grid models we extracted features using two grids: 1) Uber H3 (Resolution 10) and, 2) Geohash~(7 digits). For multiscale models we extracted features using Uber H3 grid at resolutions 6 to 12.

\textbf{Hyper-parameters tuning}. We used Hyperopt library \cite{bergstra2013making} to tune all hyper-parameters. We limited the search to 100 runs or 48 hours (Whichever comes first). Per-model search space configurations are detailed in Tables \ref{atab1}, \ref{atab2}, \ref{atab3}, \ref{atab4}, \ref{atab5}, and \ref{atab6}, respectively.

\textbf{Training}. We trained all models by minimizing the logarithmic loss for a maximum of 500 epochs. We used early stopping with $1e-3$ tolerance and patience of 10 epochs whenever possible. All models were trained using the same virtual machine equipped with 128 GB of RAM, 16 $\times$ 2.4 GHz CPUs and 2 $\times$ NVIDIA Tesla V100 GPUs. 





\textbf{MLP}. Implemented using Keras framework\footnote{\url{https://keras.io/}}, our MLP block consists of a dense layer followed by a ReLu activation layer. We set batch size to 128 and trained the network using the Adam optimizer \cite{kingma2014adam}.

\textbf{rMLP}. We applied implicit (Batch normalization \cite{ioffe2015batch} and Stochastic weight averaging \cite{izmailov2018averaging}), ensemble (Dropout~\cite{srivastava2014dropout}), structural (Skip connections \cite{he2016deep}), and data augmentation~(Gaussian noise \cite{bishop1995neural}) techniques to regularize vanilla MLPs. Our rMLP block consists of a dense layer followed by a ReLu activation layer, a Gaussian noise layer, a batch normalization layer, a dropout layer and a concatenation layer. We set batch size to 128 and trained the network using the AdamW optimizer \cite{loshchilov2017decoupled}.

\begin{table}
\centering
  \caption{$k$-NN hyper-parameters search space.}
  \label{atab1}
  \begin{tabular}{|l|ll|}
    \hline
    Hyper-parameter & Type & Range \\
    \hline
    $k$  &  Integer & [1, 33] \\
    Distance metric  &  Nominal & \{L1, L2\} \\
  \hline
\end{tabular}
\end{table}

\begin{table}
\centering
  \caption{SVM hyper-parameters search space.}
  \label{atab2}
  \begin{tabular}{|l|lll|}
    \hline
    Hyper-parameter & Type & Range & Log scale \\
    \hline
    $C$  &  \small{Continuous} & \small{[$2^{-5}$, $2^{15}$]} & \checkmark\\
    \texttt{gamma}  &  \small{Continuous} & \small{[$2^{-15}$, $2^{3}$]} & \checkmark\\
  \hline
\end{tabular}
\end{table}

\begin{table}
\centering
  \caption{XGB hyper-parameters search space.}
  \label{atab3}
  \begin{tabular}{|l|lll|}
    \hline
    Hyper-parameter & Type & Range & Log scale \\
    \hline
    \texttt{eta}  &  \small{Continuous} & \small{[$1e-3$, 1]} & \checkmark\\
    \texttt{lambda}  &  \small{Continuous} & \small{[$1e-10$, 1]} & \checkmark\\
    \texttt{alpha}  &  \small{Continuous} & \small{[$1e-10$, 1]} & \checkmark\\
    \texttt{gamma}  &  \small{Continuous} & \small{[$1e-1$, 1]} & \checkmark\\
    \texttt{num\_round}  &  \small{Integer} & \small{[1, 100]} & -\\
    \texttt{max\_depth}  &  \small{Integer} & \small{[1, 20]} & -\\
    \texttt{max\_delta\_step}  &  \small{Integer} & \small{[0, 10]} & -\\
    \texttt{min\_child\_weight}  &  \small{Continuous} & \small{[0.1, 20]} & \checkmark\\
    \texttt{subsample}  &  \small{Continuous} & \small{[0.01, 1]} & -\\
    \texttt{colsample\_bylevel}  &  \small{Continuous} & \small{[0.1, 1]} & -\\
    \texttt{colsample\_bynode}  &  \small{Continuous} & \small{[0.1, 1]} & -\\
    \texttt{colsample\_bytree}  &  \small{Continuous} & \small{[0.5, 1]} & -\\
  \hline
\end{tabular}
\end{table}

\begin{table}
\centering
  \caption{TabNet hyper-parameters search space.}
  \label{atab4}
  \begin{tabular}{|l|ll|}
    \hline
    Hyper-parameter & Type & Range \\
    \hline
    \texttt{$n_a$} & \tiny{Integer} & \tiny{\{8, 16, 24, 32, 64, 128\}} \\
    \texttt{$n_{steps}$} & \tiny{Integer} & \tiny{[3, 10]} \\
    \texttt{batch\_size} & \tiny{Integer} & \tiny{\{256, 512,1024, 2048, 4096\}} \\
    \tiny{\texttt{virtual\_batch\_size}} & \tiny{Integer} & \tiny{\{256,512,1024,2048, 4096\}} \\
    \texttt{learning\_rate} & \tiny{Continuous} & \tiny{\{0.005, 0.01, 0.02, 0.025\}} \\
    \texttt{gamma} & \tiny{Continuous} & \tiny{\{1.0, 1.2, 1.5, 2.0\}} \\
    \texttt{$\lambda_{sparse}$}  &  \tiny{Continuous} & \tiny{\{0, $10^{-6}$, $10^{-4}$, $10^{-3}$, $10^{-2}$, $10^{-1}$\}} \\
    \texttt{momentum} & \tiny{Continuous} & \tiny{\{0.6, 0.7, 0.8, 0.9, 0.95, 0.98\}} \\
  \hline
\end{tabular}
\end{table}

\begin{table}
\centering
  \caption{MLP hyper-parameters search space.}
  \label{atab5}
  \begin{tabular}{|l|lll|}
    \hline
    Hyper-parameter & Type & Range & Log scale \\
    \hline
    Hidden layers & \small{Integer} & \tiny{\{3, 6, 9\}} & - \\
    \texttt{units} & \small{Integer} & \tiny{\{128, 256, 512\}} & - \\
    \small{\texttt{learning\_rate}}  &  \small{Continuous} & \tiny{[$1e-3$, $1e-1$]} & \checkmark\\
  \hline
\end{tabular}
\end{table}

\begin{table}
\centering
  \caption{$r$MLP hyper-parameters search space.}
  \label{atab6}
  \begin{tabular}{|l|lll|}
    \hline
    Hyper-parameter & Type & Range & Log scale \\
    \hline
    \small{Hidden layers} & \small{Integer} & \small{\{3, 6, 9\}}  & -\\
    \texttt{units} & \small{Integer} & \small{\{128, 256, 512\}} & -\\
    \small{\texttt{learning\_rate}}  &  \small{Continuous} & \small{[$1e-3$, $1e-1$]} & \checkmark\\
    \small{\texttt{dropout\_rate}}  &  \small{Continuous} & \small{[0.0, 0.5]} & -\\
    \small{\texttt{weight\_decay}}  &  \small{Continuous} & \small{[$1e-6$, $1e-1$]} & \checkmark\\
    \small{\texttt{stddev}}  &  \small{Continuous} & \small{[0.0, 0.5]} & \checkmark\\
    \small{SC}  &  \small{Binary} & \small{[False, True]} & -\\
    \small{SWA}  &  \small{Binary} & \small{[False, True]} & -\\
  \hline
\end{tabular}
\end{table}

\end{document}